\documentclass[amsmath,amssymb,aps,physrev,reprint,superscriptaddress
,longbibliography
]{revtex4-2} 
\usepackage[utf8]{inputenc}     
\usepackage[english]{babel}     

\usepackage{graphicx}           
\usepackage{amsmath,amssymb}    
\usepackage{amsthm}             
\usepackage{mathtools}          
\usepackage{hyperref}           

\usepackage{siunitx}
\usepackage{multirow}
\usepackage{upgreek}

\usepackage{marvosym}
\usepackage{bm}
\usepackage{dsfont}
\usepackage{float}
\usepackage{runic}

\usepackage{braket} 
\usepackage[normalem]{ulem}

\graphicspath{{./Images/}}
\newcommand*\diff{\mathop{}\!\mathrm{d}}
\newcommand*\Diff[1]{\mathop{}\!\mathrm{d^#1}}

\usepackage{microtype}

\newcommand{\tr}{\text{tr}}

\newcommand{\proj}[1]{ \ket{#1} \kern-3pt \bra{#1} }

\begin{document}

\title{Desorption-induced decoherence of nanoparticle motion}
	
	\author{Jonas Sch\"afer}
	\affiliation{Faculty of Physics, University of Duisburg-Essen, Lotharstra\ss e 1, 47048 Duisburg, Germany}

	\author{Benjamin A. Stickler}
	\affiliation{Institute for Complex Quantum Systems and Center for Integrated Quantum Science and Technology,
		Ulm University, Albert-Einstein-Allee 11, D-89069 Ulm, Germany}

	\author{Klaus Hornberger}
	\affiliation{Faculty of Physics, University of Duisburg-Essen, Lotharstra\ss e 1, 47048 Duisburg, Germany}

	\begin{abstract}
        We derive the quantum master equation predicting how the translational and rotational dynamics of a nanoparticle is affected by the emission of 
        surface adsorbates. This is motivated by recent experiments which prepared the motion of internally hot silica particles in the deep quantum regime. 
        In the limit of a well localized nanoparticle the ro-translational dynamics can be characterized by diffusion rates in quantitative agreement with classical expectations. The theory is also suited to describe the decoherence effect of outgassing and sublimation.
	\end{abstract}
	
\maketitle

\section{Introduction}
	Owing to their exceptional isolation from environmental noise, levitated nanomechanical systems are excellent platforms for metrology and for probing new physics \cite{Gonzalez-Ballestero2021}.
	Recent advances include ultraprecise torque sensing capabilities \cite{Ahn2020} as well as optical ground state cooling of both the translational \cite{Delic2020, Magrini2021, Tebbenjohanns2021, Ranfagni2022, Kamba2022, Piotrowski2023} and librational \cite{Dania2024} degrees of freedom.
    This progress is enabled by increasing control over noise sources. The scattering of residual gas can be virtually eliminated by better vacua, and laser phase noise introduced by optical trapping or manipulation may be reduced by better optical components and less sensitive control schemes \cite{Meyer2019, Delic2020, Dania2024}.
	The currently dominant noise source in the form of photon recoil heating is unlikely to be prohibitive for the creation of macroscopic superpositions, as hybrid traps \cite{Millen2015, Fonseca2016, Conangla2020, Bykov2022, bonvin2024state, Bonvin2024} and accelerated expansion interferometric schemes 
    \cite{Romero-Isart2017, Weiss2021, Roda-Llordes2024, Neumeier2024}
    mature.

    In view of the long coherence times targeted in superposition experiments
    \cite{Bateman2014, Bose2017, Pino2018, Stickler2018, Kaltenbaek2023},
    it is thus important to consider  sources of noise that are potentially more difficult to eliminate, such as those associated with the internal makeup of the nanoparticle.
    While often approximated as rigid-bodies,
    the nanoparticles are composed of a macroscopic number of constituents
    at a finite temperature
    \cite{Millen2014, Rahman2017}
    whose interaction with the  environment and external forces can lead to additional decoherence of the center-of-mass and rotational motion. This includes the coupling to acoustic modes \cite{Gonzalez-Ballestero2020,Henkel2024} and the emission of thermal radiation \cite{Hackermuller2004, Chang2010, Agrenius2023, Neumeier2024, Schafer2024}.
    It is expedient to study how such noise affects both the center-of-mass and the rotational motion because these are in general coupled by the trapping potential. Moreover, the nonlinear quantum dynamics of rotations can offer inherent advantages  \cite{Stickler2018,Stickler2021, Gonzalez-Ballestero2021, Schrinski2022}.
	
	In this article, we present the quantum  master equation describing how the emission of constituent particles or surface contaminants decoheres the motional quantum state of the emitting bodies. 
	Specifically, for the case of adsorbate desorption, we find the Markovian master equation to be entirely characterized by the spectral particle flux density per surface element.
    The corresponding Lindblad jump operators turn out to be a product of two contributions,
    the first is a linear and angular momentum recoil due to the emission event, the second amounts to a weak orientation measurement associated with anisotropic emission patterns.
	For particles well localized in both position and orientation the dynamics become diffusive in momentum space, with an additional thermophoresis-like Hamiltonian contribution. In this limit, our predictions agree with classical results.

    We then provide the quantum description of more general outgassing processes, such as sublimation.
    It is characterized by complex emission amplitudes, which are defined in terms of the Green function associated with the dynamics of the escaping particle.
    We derive these master equations
    by casting the  decay of a metastably bound state in the body-fixed frame in terms of the M\o{}ller operators of the corresponding scattering problem. To this end, the nanoparticle motion is assumed to be much slower than each 
    emission process, and the events to be independent.

\begin{figure}
    \includegraphics[width=0.9\linewidth]{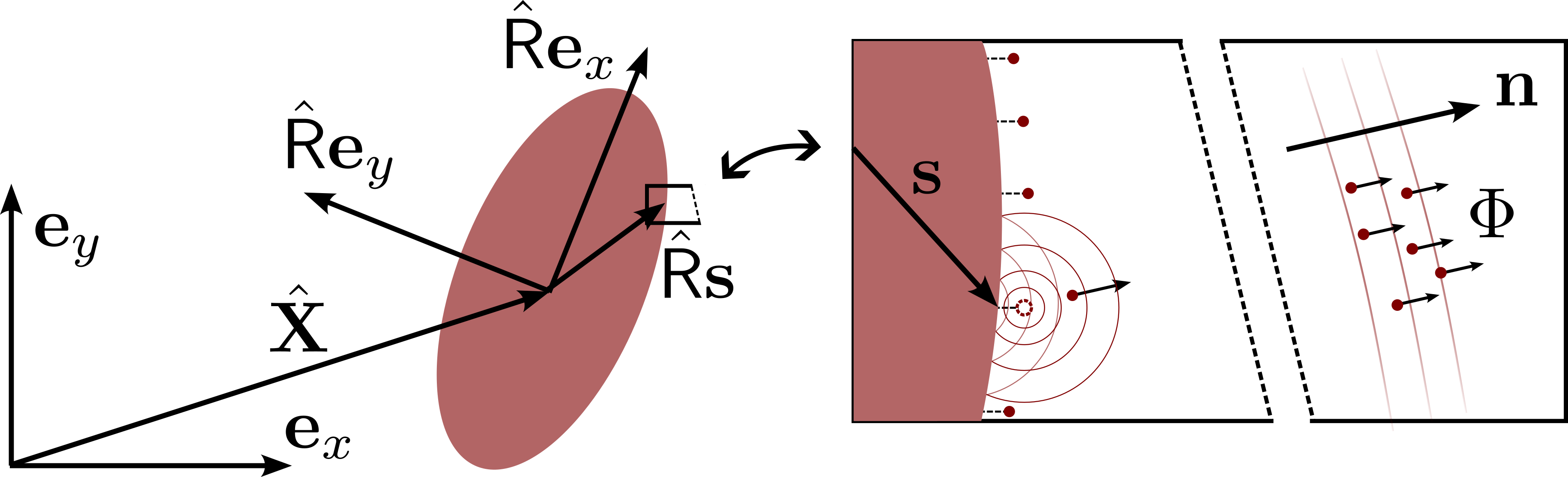}
    \caption{
    The desorption of adsorbates from the surface of a nanoparticle, whose position $\hat{\mathbf{X}}$ and orientation $\hat{\mathsf{R}}$ are quantum degrees of freedom, leads to ro-translational decoherence. It is fully characterized by the adsorbate flux density $\Phi(\mathbf{n},\mathbf{s},E)$, a function of the emission direction $\mathbf{n}$, the (body-fixed) surface point $\mathbf{s}$, and the kinetic energy $E$.
    }
    \label{fig:sketch}

\end{figure}
  	
\section{Desorption-induced decoherence}
	\label{chap:Deso}
	
	\subsection{Quantum master equation}
	\label{chap:DesoME}
    The position and orientation of the nanoparticle may be described by its  center-of-mass vector ${\mathbf{X}}$ and its orientation tensor ${\mathsf{R}}$; the latter rotates body-fixed vectors from a reference orientation to the current one.
	The associated momenta are the vectors of linear momentum ${\mathbf{P}}$ and angular  momentum ${\mathbf{J}}$.
	The corresponding quantum observables, denoted  with a hat, fulfill the commutation relations
	$[ \hat{\mathbf{X}}, \mathbf{m} \cdot \hat{\mathbf{P}} ] = i \hbar \mathbf{m} $
	and 
	$[\hat{\mathsf{R}}, \mathbf{m} \cdot \hat{\mathbf{J}} ] = i \hbar \mathbf{m} \times \hat{\mathsf{R}}$
    for any vector $\mathbf{m}$.

    The ro-translational state $\rho$ of the nanoparticle evolves according to
    \begin{align}
        \partial_t{\rho} = \dfrac{1}{i \hbar} \left[\dfrac{\hat{\mathbf{P}}^2}{2 M} + \dfrac{1}{2} \hat{\mathsf{J}} \cdot \hat{\mathsf{R}} \mathsf{I}_0^{-1} \hat{\mathsf{R}}^T \hat{\mathsf{J}} + V (\hat{\mathbf{X}}, \hat{\mathsf{R}}, t), \rho \right] + \mathcal{D} \rho
    \end{align}
    where $M$ is the nanoparticle mass and $\mathsf{I}_0$ is its tensor of inertia in the body-fixed frame, while the potential term accounts for (conservative) external forces and torques.
    In the following, we focus on the incoherent contribution to the Markovian dynamics, as described by the superoperator $\mathcal{D}$,
    resulting from the emission of nanoparticle constituents.

    We first focus on the case of desorption, where scalar particles of mass $m\ll M$ are weakly bound to the nanoparticle surface $\partial V$ at positions $\mathbf{s}$
	and are released into the vacuum over time, see Fig.~\ref{fig:sketch}.
	From now on, these emitted particles will be referred to as \emph{atoms}.
	For the nanoparticle in reference position and orientation this results in an atom flux density of
	$\Phi (\mathbf{n}, \mathbf{s}, E)$, which gives the rate of outgoing atoms per solid angle $\Diff 2 n$, per emitting surface element $\Diff 2 s$ for an atom with  energy $ E$.
	
	Assuming the nanoparticle dynamics to be much slower than the emission timescale (sudden approximation) we find that the incoherent part of the motional dynamics induced by the desorption of atoms is generated by a Lindbladian, see Sect.~\ref{chap:deriv}.
	It takes the form
	\begin{align}
        \label{eq:MEDes}
		\mathcal{D} \rho ={}&
		\int\limits_0^\infty \diff E
		\int\limits_{\partial V} \Diff2s
		\int\limits_{S^2} \Diff 2 n\,
		\nonumber\\&\times
		\big[\hat{L}(\mathbf{n}, \mathbf{s}, E) \rho \hat{L}^\dagger(\mathbf{n}, \mathbf{s}, E)-  |\hat{L}(\mathbf{n}, \mathbf{s}, E)|^2 \rho \big]
	\end{align}
	with jump operators
	\begin{align}
		\label{eq:LDes}
		\hat{L} (\mathbf{n}, \mathbf{s}, E) ={}&
		e^{- i  \mathbf{n} \cdot (\hat{\mathbf{X}} + \hat{\mathsf{R}} \mathbf{s}) p(E)/ \hbar} \Phi^{1/2} (\hat{\mathsf{R}}^T \mathbf{n}, \mathbf{s}, E)
		,
	\end{align}
	where  $\mathbf{s}\in \partial V$ are the surface positions with respect to the center of mass and $p(E)=\sqrt{2m E}$.
	Note that this equation is of Lindblad form  because $\int_{S^2} \Diff 2n \, |\hat{L}(\mathbf{n}, \mathbf{s}, E)|^2 = \int_{S^2} \Diff 2n \, \Phi(\mathbf{n}, \mathbf{s}, E)$ is not operator-valued.
    
	One can identify two conceptually distinct contributions to the jump operators. The unitary operator $e^{- i \mathbf{n} \cdot (\mathbf{X} + \mathsf{R} \mathbf{s}) p(E) / \hbar}$ imparts linear and angular momentum kicks compensating for the linear and orbital angular momentum of the outgoing atom.
    The square root of the flux, on the other hand, contributes to orientational decoherence because, for a fixed emission site,  one could obtain information on the particle orientation from the solid angle distribution of the emission pattern.
    (Fixing a site is justified since the atoms are emitted independently.)
	Consequently, this factor may contribute to decoherence even if the total spectral flux $\int_{\partial V} \Diff 2 s \, \Phi(\mathbf{n}, \mathbf{s}, E)$ is isotropic.

    \subsection{Diffusive limit}\label{chap:DiffLim}
    For large particles one is interested mainly in the
	limit of small spatio-orientational delocalizations.
	One may then take the action of the jump operators on the orientational subspace to be dominated by the orbital angular momentum recoil.
    
	As we shall see shortly, the non-unitary dynamics become diffusive in momentum space and may be characterized by a diffusion tensor.
	Following the classical treatment in  \cite{Martinetz2018}, we denote this object as a matrix of tensors, where the blocks correspond to the linear and angular degrees of freedom, respectively, and the off-diagonal blocks describe correlated diffusion of the linear and angular momentum,
	\begin{align}
\underline{\underline{\mathsf{D}_{0}}} ={}&
		\dfrac{1}{2}
		\int\limits_{0}^{\infty} \diff E
		\int\limits_{\partial V} \Diff 2 s
		\int\limits_{S^2} \Diff 2 n \,
		\Phi(\mathbf{n}, \mathbf{s}, E)
		p^2(E)
		\nonumber\\&\times
\begin{pmatrix}
	\mathbf{n} \otimes \mathbf{n} &
	\mathbf{n} \otimes (\mathbf{s} \times \mathbf{n}) \\
	(\mathbf{s} \times \mathbf{n}) \otimes \mathbf{n} &
	(\mathbf{s} \times \mathbf{n}) \otimes ( \mathbf{s} \times \mathbf{n})
\end{pmatrix}
		.
	\end{align}
    In the following, 
    multiplications of column vectors (of vectors) and matrices (of tensors) are to be understood as the corresponding operations in 6D.
    
    If the nanoparticle state $\rho$  is well-oriented around $\tilde{\mathsf{R}}$,
    and in absence of an external potential, $V=0$, the variance
    of the momentum increases linearly in time,
    \begin{align}
    \label{eq:DesMomDiff}
            \frac{\diff}{\diff t}
            \mathrm{Var}\Big[
            \begin{pmatrix} \tilde{\mathsf{R}}^T \hat{\mathbf{P}} \\ \tilde{\mathsf{R}}^T \hat{\mathbf{J}} \end{pmatrix}
            \Big]
		\approx 2
        \underline{\underline{\mathsf{D}_{0}}}
        ,
    \end{align}
    a defining characteristic of diffusion. The rotation tensors appear because $\underline{\underline{\mathsf{D}_0}}$ is defined for the reference orientation (body-fixed frame).
    (The variance of a vector is given by
    $\mathrm{Var} [\mathbf{a}] = \braket{\mathbf{a} \otimes \mathbf{a}} - \braket{\mathbf{a}} \otimes \braket{\mathbf{a}}$.)

	In addition, one gets Hamiltonian contributions corresponding to a thermophoresis-like force and torque, which can be stated succinctly as
	\begin{equation}
		\underline{\mathbf{F}_0} =
		-
		\int\limits_{0}^{\infty} \diff E
		\int\limits_{\partial V} \Diff 2 s
		\int\limits_{S^2} \Diff 2 n \,
		\Phi(\mathbf{n}, \mathbf{s}, E)
		p(E)
		\begin{pmatrix}
			\mathbf{n} \\ \mathbf{s} \times \mathbf{n}
		\end{pmatrix}
		.
	\end{equation}
    It leads to (again for $V=0$)
    a drift of the first moments of the momenta,
    \begin{align}
    \label{eq:DesMomDrift}
            \frac{\diff}{\diff t}
            \Big\langle
        \begin{pmatrix} \tilde{\mathsf{R}}^T \hat{\mathbf{P}} \\ \tilde{\mathsf{R}}^T \hat{\mathbf{J}} \end{pmatrix}
        \Big\rangle
        \approx{}&
        \underline{\mathbf{F}_0}
        .
    \end{align}
    
    To obtain the diffusive limit
    consider Eq.~(\ref{eq:MEDes}) for a state $\rho$  well-localized around $\tilde{\mathsf{R}}$.
The orientation tensor operator $\hat{\mathsf{R}}$ can then be approximated by means of the vector operator $\hat{\mathbf{w}} = \sum_{j=1}^3 \mathbf{e}_j \times \log (\tilde{\mathsf{R}}^T \hat{\mathsf{R}}) \mathbf{e}_j /2$, with orthonormal basis vectors $(\mathbf{e}_1,\mathbf{e}_2,\mathbf{e}_3)$.
	The length and direction of $\hat{\mathbf{w}}$ give the operator-valued rotation angle and axis of rotation relative to ${\tilde{\mathsf{R}}}$.
	For small relative angles we have to lowest order $\tilde{\mathsf{R}} \hat{\mathbf{w}} \approx \sum_{j=1}^3 ( \tilde{\mathsf{R}} \mathbf{e}_{j}) \times (\hat{\mathsf{R}} \mathbf{e}_{j})/2$,
	and in turn
	$\hat{\mathsf{R}} \mathbf{e}_j \approx \tilde{\mathsf{R}} [\mathbf{e}_j + \hat{\mathbf{w}} \times \mathbf{e}_j]$.
	This allows us to approximate the master equation (\ref{eq:MEDes})
    without introducing coordinates,
    \begin{align}
		\label{eq:MEDesoDiff}
		\mathcal{D} \rho \approx{}&
		\dfrac{1}{i \hbar} \bigg[- \begin{pmatrix} \tilde{\mathsf{R}}^T \hat{\mathbf{X}} \\  \hat{\mathbf{w}} \end{pmatrix}
		\cdot 
        \underline{\mathbf{F}_0}
        ,
        \rho \bigg]
		\nonumber\\&
		+ \dfrac{2}{\hbar^2}
		\Bigg[
		\begin{pmatrix} \tilde{\mathsf{R}}^T \hat{\mathbf{X}} \\  \hat{\mathbf{w}} \end{pmatrix}
		\cdot
        \underline{\underline{\mathsf{D}_{0}}} \,
        \rho
		\begin{pmatrix} \tilde{\mathsf{R}}^T \hat{\mathbf{X}} \\\hat{\mathbf{w}} \end{pmatrix}
		\nonumber\\&
        - \dfrac{1}{2}
		\Big\{
		\begin{pmatrix} \tilde{\mathsf{R}}^T \hat{\mathbf{X}} \\\hat{\mathbf{w}} \end{pmatrix}
		\cdot
        \underline{\underline{\mathsf{D}_{0}}}
		\begin{pmatrix} \tilde{\mathsf{R}}^T \hat{\mathbf{X}} \\\hat{\mathbf{w}} \end{pmatrix}
		, \rho \Big\}
		\Bigg]
		.
    \end{align}
    From this one arrives at Eqs.~(\ref{eq:DesMomDiff}) and (\ref{eq:DesMomDrift})
    by keeping only the lowest order of the commutator $[ \hat{\mathsf{R} } \mathbf{e}_j, \mathbf{e}_\ell \cdot \hat{\mathbf{J}} ] \approx i \hbar \mathbf{e}_\ell \times (\tilde{\mathsf{R}} \mathbf{e}_{j})$, i.e. neglecting higher orders of $ \hat{\mathsf{R}} - \tilde{\mathsf{R}}$.

	If the flux density is inversion symmetric,
    $\Phi(-\mathbf{s} , - \mathbf{n}, E) = \Phi (\mathbf{n}, \mathbf{s}, E)$, the linear and angular degrees of freedom decouple and the force vanishes. A vanishing torque however requires an additional symmetry.
	
	For certain forms of $\Phi (\mathbf{n}, \mathbf{s}, E)$ and bodies of high symmetry, the integrals can be solved analytically.
	As an example, consider a flux density with cosine-shaped angular distribution, $\Phi(\mathbf{n}, \mathbf{s}, E) =  \mathbf{n}\cdot\mathbf{n}_\mathrm{s} \,\Theta (\mathbf{n}\cdot\mathbf{n}_\mathrm{s})\Phi_0 (E)$, where $\mathbf{n}_\mathrm{s} \equiv \mathbf{n}_\mathrm{s} (\mathbf{s})$ is the local surface normal and $\Theta$ is the Heaviside step function.
	One can then directly carry out the solid angle integral. The force and torque vanish in this case, as may be seen by invoking the divergence theorem, while the diffusion tensor reads as
	\begin{align}
		\underline{\underline{\mathsf{D}_{0}}}
                    ={}&
		\dfrac{\pi}{8}
		\int\limits_{0}^{\infty} \diff E \,
		\Phi_0 (E)
		p^2 (E)
		\int\limits_{\partial V} \Diff 2 s \,
		\nonumber\\&\times
        \bigg[
             \begin{pmatrix}
			\mathds{1} & - [\mathbf{s}]_\times \\
                [\mathbf{s}]_\times & -[\mathbf{s}]_\times^2
              \end{pmatrix}
              +
              \begin{pmatrix}
			\mathbf{n}_\mathrm{s} \\
                \mathbf{s} \times \mathbf{n}_\mathrm{s}
              \end{pmatrix}
              \otimes
              \begin{pmatrix}
			\mathbf{n}_\mathrm{s} \\
                \mathbf{s} \times \mathbf{n}_\mathrm{s}
              \end{pmatrix}
                    \bigg]
                    .
	\end{align}
    Here, $[\mathbf{s}]_{\times}$ denotes the skew-symmetric tensor representation of the cross product with $\mathbf{s}$, i.e. $[\mathbf{s}]_{\times} \mathbf{a} = \mathbf{s} \times \mathbf{a}$ for any vector $\mathbf{a}$.

	As a consistency check, we compare our results to those obtained by a classical model.
	Considering the flux  obtained in the classical description of the desorbtive part of diffusive gas-scattering \cite{Martinetz2018}
    both the resulting diffusion tensor
    and the thermophoresis-like force and torque
    agree
    \footnote{See \cite{Martinetz2018}: For the diffusion tensor consider the $\gamma_\mathrm{s}^2$ contributions in Eq.~(41) as well as Table II; for the force see Eq.~(38).}.
	
	\section{General outgassing master equation}
    \label{chap:GenEm}

    We now consider general outgassing processes, where the source points of the ejected atoms are no longer confined to the particle surface.
    In this case, the potential imposed on the escaping atom by the delocalized nanoparticle imprints additional operator-valued phases which appear in the jump operators.
	The outgoing particle flux distribution then no longer suffices to determine the master equation, as the former does not contain phase information.

    For general emission processes,  the master equation is characterized by complex-valued amplitudes of emission. To define the latter, we take the  nanoparticle to be in reference position and orientation, and consider the retarded Green function $G^+ ( \mathbf{r},\mathbf{s}, E)$ of the Schrödinger equation for the escaping atom.
    It describes the wavefunction at position $\mathbf{r}$ due to a source at position $\mathbf{s}$ at fixed energy $E$, and yields the amplitude of emission
    in the limit
    \begin{equation}\label{eq:Adef}
        A^+ (\mathbf{n}, \mathbf{s}, E) = \lim_{r \to \infty} - r e^{- i p(E) r/\hbar} G^+ ( r\mathbf{n},\mathbf{s}, E)
        .
    \end{equation}
	This object may be understood as the  analogue to the scattering amplitude for an emission problem. 
    
    To specify the general master equation, all that is needed in addition is the total emission rate due to a source at point $\mathbf{s}$, denoted by $\Gamma (\mathbf{s}, E) $. As shown in Sect.~\ref{chap:deriv}, the jump operators of the master equation then read  
	\begin{align}\label{eq:LGen}
		\hat{L} (\mathbf{n}, \mathbf{s}, E) ={}&
		\sqrt{\Gamma(\mathbf{s}, E)}
		\bigg[ \int\limits_{S^2} \Diff2n'  \, | A^+ (\mathbf{n}', \mathbf{s} , E)|^2 \bigg]^{-1/2}
		\nonumber\\&\times
		e^{- i p(E) \mathbf{n} \cdot \hat{\mathbf{X}} /\hbar} A^+ (\hat{\mathsf{R}}^T \mathbf{n}, \mathbf{s}, E)
        .
	\end{align}
    The master equation takes the form of (\ref{eq:MEDes}), with the surface integral  $\int_{\partial V} \Diff2s$ replaced by the  volume integral $\int_{ V} \Diff3s$, see (\ref{eq:MEGen}).
    
    Furthermore, the jump operators give the outgoing particle flux density,
	\begin{equation}
		\label{eq:GenFlux}
		\Phi (\hat{\mathsf{R}}^T \mathbf{n} , \mathbf{s}, E) =  |\hat{L}(\mathbf{n}, \mathbf{s}, E)|^2
	\end{equation} 
    and they are normalized as $\int_{S^2} \Diff 2 n \, |\hat{L}(\mathbf{n}, \mathbf{s}, E)|^2 = \Gamma (\mathbf{s}, E)$.

	\section{Derivation of the master equation}
	\label{chap:deriv}	
	We proceed to derive the general Markovian outgassing master equation describing the impact of particle emission in terms of the Lindblad operators (\ref{eq:LGen}). This is done by combining the description  of a metastably bound atom in the `weak-coupling limit' with the assumption of strongly separated motional and emission timescales (`{sudden approximation}'), and utilizing the {resolvent formalism} for the escaping atom in presence of the nanoparticle potential.
    This line of reasoning was already employed in the context of the emission of thermal radiation from dielectric particles \cite{Schafer2024}. Here we give a more detailed and self-contained account of the derivation.
    
\subsection{Emission of metastably bound atoms}
	
	We fix the nanoparticle in reference position and orientation for the time being.
    To start with, consider an atom metastably bound in the state $\ket{\varphi_0}$ at energy $E_0$. 
    Ultimately, we are interested in the unbound part of the atomic motion, described by a wave function subject to the nanoparticle potential.
    To model the decay one may introduce a two-level system, which controls whether the atom feels a trapping  potential, together with a coupling term inducing transitions with a rate $g$,
	\begin{align}
		H_\mathrm{tot} ={}&
		E_0 \proj{\varphi_0}\otimes \proj{\uparrow} + H\otimes \proj{\downarrow}
		\nonumber\\&
		+ g \hbar
		\proj{\varphi_0}\otimes (\ket{\uparrow}\bra{\downarrow}+\ket{\downarrow}\bra{\uparrow})
		.
	\end{align}
	Here, $H = p^2/2m + V$ is the Hamilton operator of the escaping atom (assumed to have no bound states).
	To solve the time evolution we decompose the state as
	\begin{equation}\label{13}
		\ket{\Psi(t)} ={}
		e^{- i E_0 t /\hbar} b(t) \ket{\varphi_0} \ket{\uparrow} + \int\limits_{0}^{\infty} \diff E \, c_E (t) e^{- i E t /\hbar} \ket{E}\ket{\downarrow}
		,
	\end{equation}
        taking the eigenstates $\ket{E}$ of $H$ to be nondegenerate for the time being.
	Inserting (\ref{13}) into the Schrödinger equation one finds the coefficients to obey
	\begin{align}
		\dot{b} (t) ={}&
		- i g \int\limits_0^\infty \diff E \, e^{- i (E- E_0) t /\hbar} \braket{\varphi_0 | E} c_E (t)
        \\
		\dot{c}_{E}(t) ={}&
		- i g e^{i (E- E_0) t /\hbar} \braket{E | \varphi_0} b(t)
		.
	\end{align}
	We formally integrate the coefficient of the unbound part, insert it into the evolution of the bound part and perform a Born-Markov-like approximation, by replacing $b(t') \to b(t)$ and extending the temporal integration to include the infinite past,
	\begin{align}
		\dot{b} (t) ={}&
		- g^2 \int\limits_0^\infty \diff E \, |\braket{E| \varphi_0} |^2 \int\limits_0^t \diff t' \, e^{- i (E - E_0) (t - t')/\hbar} b(t')
        \nonumber
		\\
		\approx{}&
		- b(t) \hbar g^2 \int\limits_0^\infty \diff E \, |\braket{E_0| \varphi_0} |^2
		\nonumber\\&\times
		\left[ \pi \delta (E- E_0) - i \mathcal{P} \dfrac{1}{E - E_0}\right]
		.
	\end{align}
	  Neglecting the energy renormalization due to the principal-value, the solution for the initial condition $b(0) = 1$  reads $b(t) = e^{- \Gamma_0/2 t}$, with the probability decay rate 
\begin{align}
\label{eq:DefGam0}
     \Gamma_0 = 2 \pi \hbar g^2 |\braket{E_0 | \varphi_0}|^2 .
\end{align}
      	Reinsertion into the formal solution of the escaping part of the wave function $\ket{\psi_\downarrow} = \braket{\downarrow | \Psi}$ gives, for $\Gamma_0 t \ll1$,
	\begin{align}
		\ket{ \psi_\downarrow (t)} ={}&
		- 
		e^{- i E_0 t /\hbar } \hbar g \int\limits_{0}^{\infty} \diff E \,
		\big(
		1
		+ e^{- i (E-E_0) t /\hbar}
		\big)
		\nonumber\\&\times
		\dfrac{\proj{E}}{E_0 - i \hbar \Gamma_0 /2 - E } \ket{\varphi_0}
		.
	\end{align}
	In the first term of the integrand one may  identify the resolvent operator $G(z) = [z - H]^{-1}$ with $z=E_0 - i \hbar \Gamma_0/2$.
	To deal with the second term, we again consider the asymptotics of large $t$. In this case, the off-resonant contributions
	oscillate arbitrarily fast and average out, given that the remaining integrand is a smooth integrable function of $E$. This effectively allows one to replace the energy projector by $\proj{E_0}$; the same argument allows extending the lower integration boundary to negative infinity. The remaining integral can then be solved using contour integration,
    $\int_{-\infty}^\infty \diff E \, e^{- i E t/ \hbar} / (E + i \hbar \Gamma_0/2) = e^{- \Gamma_0 t/2} 2 \pi i$,
    and the projector may also be expressed using resolvent operators since $ - 2 \pi i \proj{E}=G(E+i 0^+)-G(E-i 0^+)$.
	Since $\Gamma_0t\ll1$ this leaves us with the approximate solution
	\begin{align}\label{eq:psidown}
		\ket{\psi_\downarrow (t)}
		\sim
		\hbar g e^{- i E_0 t /\hbar} G (E_0 + i \hbar \Gamma_0/2) \ket{\varphi_0}
	\end{align}
	as $t\to\infty$. It describes the arbitrarily slow leaking of a metastably bound state acting as a source term in the Schrödinger equation for the escaping part of the wavefunction, 
    \begin{align}
    (i\hbar\partial_t - {H})\ket{\psi_\downarrow} = \hbar g \exp(- i E_0 t /\hbar)\ket{\varphi_0}.
    \end{align}
	We note that the asymptotic solution (\ref{eq:psidown}) has unit norm since
    \begin{align}
        \lim_{\Gamma_0\to0}
        \dfrac{\Gamma_0}{\pi}
        G(E-i \hbar \Gamma_0 /2) G(E+ i \hbar \Gamma_0/2) 
        =
         \proj{E},
    \end{align}
    as follows from 
    $ (z'-z) G(z) G(z') = G(z) - G(z')$ (first resolvent identity)
    and the Sokhotski–Plemelj formula.
	
	\subsection{Impact of emission events on the motional degrees of freedom}
    Let us now consider how the emission process acts back on the motional state of the emitting particle, assuming the latter to be static on the emission timescale.
    (In this approximation the kinetic energies of the atom in the body-fixed frame and the lab frame coincide.)
    The active transformation from the reference to the actual particle position and orientation is described by the unitary operator $\hat{D} \equiv D (\hat{\mathbf{X}}, \hat{\mathsf{R}})$, acting on the total nanoparticle-atom Hilbert space.
    Introducing the position-orientation basis as the common eigenvectors of $\hat{\mathbf{X}}$ and $\hat{\mathsf{R}}$, denoted $\ket{\mathbf{X}, \mathsf{R}} \equiv \ket{\mathbf{X}} \ket{\mathsf{R}}$, the  transformation takes the form $ \hat{D} \ket{\mathbf{r}} \ket{\mathbf{X}, \mathsf{R} } = \ket{\mathsf{R} \mathbf{r} + \mathbf{X}} \ket{\mathbf{X}, \mathsf{R}} $.

    The previous calculation is easily adapted to the transformed problem,
    i.e.\ with $V \to \hat{D} V \hat{D}^\dagger$ and $\ket{\varphi_0} \to \hat{D} \ket{\varphi_0}$, which shows that the state of the escaping atom  is also transformed as
    $\ket{\psi_{\downarrow}} \to \hat{D} \ket{\psi_{\downarrow}}$.    Expanding $\hat{D}$ in a position-orientation basis demonstrates that the escaping atom state is highly correlated with the nanoparticle state of motion. 
	One may thus obtain the back-action on the reduced nanoparticle state conditioned on emission by tracing out the atomic state,
	\begin{equation}\label{eq:FormAtomTr}
		\rho' = \tr_{\mathrm{atom}} \big[ \hat{D} \proj{\psi_\downarrow} \otimes \rho \hat{D}^\dagger
		\big]
		.
	\end{equation}
	
	To evaluate the trace we use the second resolvent identity, $G = G_0 (\mathds{1} + V G) = (\mathds{1} + G V) G_0$, and identify the M\o{}ller operators $\Omega_{\pm} = \lim_{t \to \mp \infty} e^{ i H t / \hbar} e^{ - i H_0 t / \hbar}$ where $H_0 = H - V$.
	The latter map asymptotically incoming and outgoing states to the corresponding scattering states, and they are unitary since $H$ has no bound states \cite{Newton1982}. 
	The time limit in the definition can be shown to be equivalent to the so-called
    adiabatic switching limit \cite{Dollard1966, Newton1982}
	\begin{equation}
		\Omega_{\pm} =
		\lim\limits_{\epsilon \to 0^+} \pm \epsilon \int\limits_{\mp \infty}^0 \diff t \,
		e^{\pm \epsilon t} e^{i H t} e^{- i H_0 t}
		.
	\end{equation}
	It will be useful to consider the action of the M\o{}ller operators on an eigenstate $\ket{E}_0$ of $H_0$,
	\begin{align}
		\Omega_{\pm} \ket{E}_0 ={}&
		\lim\limits_{\epsilon \to 0^+} \pm \epsilon \int\limits_{\mp \infty}^0 \diff t \,
		e^{\pm \epsilon t} e^{i H t} e^{- i E t} \ket{E}_0
		\nonumber\\={}&
		\lim\limits_{\epsilon \to 0^+} \pm i \epsilon G(E \pm i \epsilon) \ket{E}_0
            \nonumber\\={}&
            [\mathds{1} + G (E \pm i 0^+) V ]\ket{E}_0
		\label{eq:AdSwMoll}
		,
	\end{align}
    where the final line follows from the second resolvent identity.

	Returning to Eq.~(\ref{eq:FormAtomTr}), we perform the trace in momentum basis and use that $D^\dagger (\hat{\mathbf{X}}, \hat{\mathsf{R}}) \ket{\mathbf{p}} = e^{i \mathbf{p} \cdot \hat{\mathbf{X}} / \hbar} \ket{ \hat{\mathsf{R}}^T \mathbf{p}}$. Here,  the operator-valued atomic momentum state is
    to be understood in terms of an expansion in the position-orientation basis of the nanoparticle.
	Together with the second resolvent identity we arrive at
	\begin{align}
		\rho'
		={}&
				\lim\limits_{\Gamma_0 \to 0^+}
				\hbar^2 g^2
				\int\limits_{\mathds{R}^3} \Diff 3 p \,
				e^{- i \mathbf{p} \cdot \hat{\mathbf{X}}/\hbar}
				\bra{\hat{\mathsf{R}}^T \mathbf{p}} G_0(E_0 + i \hbar \Gamma_0 /2)
				\nonumber\\&\times
				[\mathds{1} + V G (E_0 + i \hbar \Gamma_0 /2)]
				\proj{\varphi_0} \otimes \rho \,
				\nonumber\\&\times
				[\mathds{1} + V G (E_0 + i \hbar \Gamma_0 /2)]^\dagger G_0^\dagger (E_0 + i \hbar \Gamma_0 /2)
				\nonumber\\&\times
				\ket{\hat{\mathsf{R}}^T \mathbf{p}}
				e^{i \mathbf{p} \cdot \hat{\mathbf{X}}/\hbar}
.
\end{align}
Since $G_0$ is diagonal in momentum this yields a nascent delta function, which enforces the on-shell momentum $p_0 = \sqrt{2 m E_0}$, and this allows one to identify the M\o{}ller operators for outgoing states based on (\ref{eq:AdSwMoll}),
\begin{align}
\rho'
				={}&
		\dfrac{1}{|\braket{E_0 | \varphi_0}|^2}
		\lim\limits_{\Gamma_0 \to 0^+}
		\int\limits_{S^2} \Diff 2 n \int\limits_0^\infty \diff p \,
		\dfrac{\hbar \Gamma_0 /2 \pi}{(E_p - E_0)^2 + \hbar^2 \Gamma_0^2 /4}
		\nonumber\\&\times
		\bra{ p \mathbf{n}} \hat{D} [\mathds{1} + V G (E_0 + i \hbar \Gamma_0 /2)]
		\proj{\varphi_0} \otimes \rho
		\nonumber\\&\times
		[\mathds{1} + V G (E_0 + i \hbar \Gamma_0 /2)]^\dagger \hat{D}^\dagger \ket{p \mathbf{n} }
		\nonumber\\
		={}& 
		\dfrac{m p_0}{|\braket{E_0 | \varphi_0}|^2}
		\int\limits_{S^2} \Diff 2 n \,
		\bra{ p_0 \mathbf{n}} \hat{D} \Omega_{-}^\dagger
        \ket{\varphi_0} \rho \bra{\varphi_0}
		\Omega_{-} \hat{D}^\dagger \ket{p_0 \mathbf{n} }
		.
        \label{eq:rhoprime}
	\end{align}
	As a consistency check, we next confirm that $\rho'$ is normalized.
    From the intertwining relation $\Omega_{\pm} H_0 = H \Omega_{\pm} $ we have
	\begin{align}
        \label{eq:OrIntToE}
		\proj{E_0}
		={}&
		\delta(E_0 - H)
		=
		\int\limits_{\mathds{R}^3} \Diff 3 p \, \, \Omega_- \proj{\mathbf{p}} \Omega_{-}^\dagger \delta (E_0 - \dfrac{p^2}{2m})
		\nonumber\\={}&
		m p_0\int\limits_{S^2} \Diff2n  \,\Omega_-\proj{p_0 \mathbf{n}}\Omega_-^\dagger
        .
	\end{align}
    Evaluating the trace in position-orientation basis then gives
	\begin{align}
		\tr [\rho']
		={}&
		\dfrac{m p_0}{|\braket{E_0 | \varphi_0}|^2}
		\int\limits_{\mathds{R}^3} \Diff 3 X
		\!\! \int\limits_{\mathrm{SO}(3)} \!\! \diff \mu (\mathsf{R}) \,
		\braket{\mathbf{X}, \mathsf{R} | \rho | \mathbf{X}, \mathsf{R}}
		\nonumber\\&\times
		\int\limits_{S^2} \Diff 2 n \,
		\bra{\varphi_0} \Omega_- \proj{p_0 \mathsf{R} \mathbf{n} }
		\Omega_-^\dagger \ket{\varphi_0}
		\nonumber\\
		={}&
		\dfrac{1}{|\braket{E_0 | \varphi_0}|^2}
		\tr [\rho]
		\braket{\varphi_0 | E_0 }  \braket{E_0 | \varphi_0} = 1
		.
	\end{align}
    The integration over rotation tensors
    covers all possible particle orientations such that $\int_{\mathrm{SO(3)}} \diff \mu(\mathsf{R}) \, \proj{\mathsf{R}} $ is a resolution of identity on the orientational subspace.

	\subsection{Dynamic equation for particle emission}
To obtain the incoherent part $\mathcal{D}\rho$ of the nanoparticle equation of motion,
we consider a time step $\diff t$ small compared to both the free nanoparticle evolution and the lifetime of a metastably bound atom, but much greater than the timescale of the emission process of an escaping atom. Given the nanoparticle state $\rho(t)$, the time-evolved one is a mixture of the transformed state (\ref{eq:rhoprime}), weighted with the transmission probability $\Gamma_0 \diff t\ll 1$, and the unperturbed state, weighted with the probability for no-emission having taken place, i.e.\ $\rho(t+\diff t)=\Gamma_0 \diff t\rho'(t)+(1- \Gamma_0 \diff t) \rho(t)$. The differential quotient then takes the form
	\begin{align}\label{eq:MeGenSgCh}
		\mathcal{D} \rho ={}&
		\Gamma_0
		\Big[
		\dfrac{m p_0}{|\braket{E_0 | \varphi_0}|^2}
		\int\limits_{S^2} \Diff 2 n \,
		\bra{ p_0 \mathbf{n}} \hat{D} \Omega_{-}^\dagger
		\ket{\varphi_0} \rho \, \bra{\varphi_0}
		\nonumber\\&\times
		\Omega_{-} \hat{D}^\dagger \ket{p_0 \mathbf{n} }
		- \rho
		\Big]
		.
	\end{align}
    It follows from (\ref{eq:OrIntToE}) that this equation is of Lindblad form.
    
For simplicity, we take the  bound states $\ket{\varphi_0} $ to be well localized, such that we can approximate them in terms of position eigenstates $\ket{\mathbf{s}}$. 
Equation (\ref{eq:MeGenSgCh}) can then be written as
\begin{align}\label{eq:MEsinglesite}
    \mathcal{D} \rho = \int\limits_{S^2} \Diff2n\, [\hat{L} (\mathbf{n} ) \rho \hat{L}^\dagger (\mathbf{n} ) - \{\hat{L}^\dagger (\mathbf{n} ) \hat{L} (\mathbf{n} ),\rho\}/2]    
\end{align}
with with jump operators
    \begin{align}\label{eq:defL}
    \hat{L} (\mathbf{n}) ={}&
    \dfrac{\sqrt{\Gamma_0 m p_0}}{|\braket{E_0 | \mathbf{s}}|}
    \bra{ p_0 \mathbf{n}} \hat{D} \Omega_{-}^\dagger
		\ket{\mathbf{s}}
        \nonumber\\={}&
    \sqrt{\Gamma_0}
    \dfrac{\bra{ p_0 \mathbf{n}} \hat{D} \Omega_{-}^\dagger
		\ket{\mathbf{s}}}{\sqrt{\int_{S^2} \Diff 2n '|\braket{p_0 \mathbf{n}'| \Omega_-^\dagger | \mathbf{s}}|^2}}
        .
    \end{align}
The matrix elements
in (\ref{eq:defL}) can be related to a limit of the Green function  $G^{\pm} (\mathbf{r},\mathbf{r}'; E_0)=\bra{\mathbf{r}}G(E_0 \pm i 0^+)\ket{\mathbf{r}'}$ of the position-space Schrödinger equation for the atom.
Specifically, they are given in terms of the amplitude of emission $A^+$, defined as
	\begin{equation}
    \label{eq:EmAmDef}
		A^\pm (\mathbf{n}, \mathbf{s}, E_0) = - \lim\limits_{r \to \infty} r e^{\mp i p_0 r /\hbar} G^{\pm}(r \mathbf{n}, \mathbf{s}, E_0)
		.
	\end{equation} 
    ($A^-$ may be understood as the corresponding amplitude of absorption.)
    
    To show this connection,
    we first consider
    a Fourier-like solid-angle integral over a smooth function $g(\mathbf{n})$, 
	\begin{align}
        \label{eq:SphHarComp}
		I (\mathbf{n}; \lambda) ={}&
		\int\limits_{S^2} \Diff 2 n' \, e^{\pm i \lambda \mathbf{n}' \cdot \mathbf{n}} g(\mathbf{n}')
		\\
		={}&
		4 \pi \int\limits_{S^2} \Diff 2  n' \sum\limits_{\ell,m} i^\ell j_\ell (\lambda) Y_{\ell m}^* (\mathbf{n}')
		Y_{\ell m} (\pm \mathbf{n})
		g(\mathbf{n}'),
        \nonumber
		\end{align}
        where we expanded the exponential in terms of spherical harmonics. One may use the asymptotic form of the spherical Bessel functions as well as properties of the spherical harmonics (see $1.17.25$ and $14.30.7$ in Ref.~\cite{DLMF}) to obtain that
		\begin{align}
		I (\mathbf{n}; \lambda)
		\sim{}&
		- 2 \pi i \int\limits_{S^2} \Diff 2  n' \sum\limits_{\ell,m} i^\ell 
		[e^{i \lambda} - (-1)^\ell e^{- i \lambda}]
		\nonumber\\&\times
		Y_{\ell m}^* (\mathbf{n}')
		Y_{\ell m} (\pm \mathbf{n})
		g(\mathbf{n}')
        \nonumber
		\\
		={}&
		\mp i \dfrac{2 \pi}{\lambda} \big[ e^{ \pm i \lambda} g(\mathbf{n}) - e^{ \mp i \lambda} g(-\mathbf{n}) \big]
	\end{align}
as $\lambda\to\infty$. With this we can now
    tackle
    the position-momentum matrix elements of the M\o{}ller operators by by inserting a position basis into the adiabatic switching formulation (\ref{eq:AdSwMoll})
	\begin{align}\label{34}
		\braket{p_0 \mathbf{n}| \Omega_{\pm} ^\dagger | \mathbf{s}} ={}&
		\lim\limits_{\epsilon \to 0^+}
		\mp i \epsilon \int\Diff3 r\, \braket{p_0 \mathbf{n} | \mathbf{r}} \braket{\mathbf{r}| G(E_0 \mp i \epsilon) |\mathbf{s} }.
        \end{align}
    At his point, we incorporate the physical assumption that the emitted atom does not get trapped by the nanoparticle potential, by taking the Green function in (\ref{34}) to be locally integrable. 
    Since we consider the limit $\epsilon \to 0$, bounded domains do not contribute in the spatial integration, allowing us to introduce a lower cut-off radius $R$,
        \begin{align}
		\braket{p_0 \mathbf{n}| \Omega_{\pm} ^\dagger | \mathbf{s}} ={}&
		\lim\limits_{\epsilon \to 0^+}
		\mp i \epsilon [2 \pi \hbar]^{-3/2} \int\limits_{R}^{\infty}\diff r \, r^2 \int\Diff 2n'
		\nonumber\\&\times
		e^{- i p_0 r \mathbf{n}' \cdot \mathbf{n}/\hbar} G(r \mathbf{n}', \mathbf{s}; E_0 \mp i \epsilon).
        \end{align}
        The cut-off can be chosen arbitrarily large, so that we may use Eq.~(\ref{eq:SphHarComp}) as well as the asymptotic form of the Green function (as implied by Eq.~(\ref{eq:EmAmDef})).
        Because $\epsilon$ is still finite at this point, the on-shell momentum takes the form
        $p(E_0 \pm i \epsilon) = \sqrt{2m (E_0 \pm i \epsilon)} \simeq p_0 \pm i \epsilon m/p_0$, so that
        \begin{align}
        \braket{p_0 \mathbf{n}| \Omega_{\pm} ^\dagger | \mathbf{s}}
		={}&
		\lim\limits_{\epsilon \to 0^+}
		\pm \epsilon \dfrac{2 \pi \hbar}{p_0} [2 \pi \hbar]^{-3/2} \int\limits_{R}^{\infty}\diff r \,
		e^{-   \epsilon m r /p_0\hbar }
		\nonumber\\&\times
		\Big[- e^{ - i p_0 r (1 \pm 1)/\hbar} A^{\mp} (\mathbf{n}, \mathbf{s}, E_0)
		\nonumber\\&
        + e^{ i p_0 r (1 \mp 1)/\hbar} A^{\mp} (-\mathbf{n}, \mathbf{s}, E_0)\Big]
.
\end{align}
The radial integration and the limit $\epsilon\to0^+$ can now be computed,
\begin{align}\label{eq:RadInt}
		\lim\limits_{\epsilon \to 0^+}
        \epsilon \int\limits_R^\infty \diff r \, e^{\pm i b r} e^{- a \epsilon r}
		=
		\begin{cases}
			0 &, b\neq 0 \\
			a^{-1} &, b=0
		\end{cases}
		.
	\end{align}
It follows that the matrix elements of the M\o{}ller operators finally evaluate to
\begin{align}\label{eq:MoMatEl}
\braket{p \mathbf{n}| \Omega_{\pm} ^\dagger | \mathbf{s}}
		={}&
		4 \pi [2 \pi \hbar]^{-3/2} \dfrac{\hbar^2}{2 m} A^{\mp} (\mp \mathbf{n}, \mathbf{s}, E_0)
		.
	\end{align}
    
The jump operator (\ref{eq:defL}) is thus given by
	\begin{equation}\label{L39}
		\hat{L} (\mathbf{n}) = 
		\sqrt{\Gamma_0} \,
        e^{- i p_0 \mathbf{n} \cdot \hat{\mathbf{X}}/ \hbar}
		\dfrac{A^{+} (\hat{\mathsf{R}}^T \mathbf{n}, \mathbf{s}, E_0)}{\sqrt{\int_{S^2} \Diff 2 n' \, | A^+ (\mathbf{n}', \mathbf{s},E_0) |^2 } }
		.
	\end{equation}
    To illustrate its action,
    let us consider the idealized case of a transparent emitter,  where the atom does not interact with the nanoparticle after release. The retarded Green function is then given by
	\begin{equation}
		G_0^+ (\mathbf{r}, \mathbf{s}; E_0) =
		- \dfrac{2m}{\hbar^2}
		\dfrac{1}{4 \pi} \dfrac{e^{i p_0 | \mathbf{r} - \mathbf{s} |/\hbar }}{ | \mathbf{r}- \mathbf{s} | },
	\end{equation}
	so that it is  straightforward to calculate the amplitude of emission (\ref{eq:EmAmDef}),
    \begin{equation}
        \label{eq:A0def}
        A_0^+ (\hat{\mathsf{R}}^T \mathbf{n}, \mathbf{s}, E_0) = \frac{m}{2 \pi \hbar^2} e^{- i p_0 \mathbf{n} \cdot \hat{\mathsf{R}} \mathbf{s} / \hbar } 
        ,
    \end{equation}
    by using $| r \mathbf{n} - \mathbf{s}| = r - \mathbf{n} \cdot \mathbf{s} + \mathcal{O} (r^{-1})$.
       
    The appearance of the operator-valued complex phase factor can be understood in two complementary ways. Consider an emission into the lab-fixed direction $\mathbf{n}$. 
    On the one hand, for two distinct particle orientations $\mathsf{R}$ and $\mathsf{R}'$
    the associated paths traversed by the atom differ by $\mathbf{n} \cdot (\mathsf{R} - \mathsf{R}')\mathbf{s}$, and therefore the accumulated phase shifts may contain information on the particle orientation. 
    On the other hand, the exponential in (\ref{eq:A0def}) effects an angular momentum kick, $e^{- i p_0 \mathbf{n} \cdot \hat{\mathsf{R}} \mathbf{s} / \hbar} \hat{\mathbf{J}} e^{i p_0 \mathbf{n} \cdot \hat{\mathsf{R}} \mathbf{s} / \hbar} = \hat{\mathbf{J}} - (\hat{\mathsf{R}} \mathbf{s}) \times p_0 \mathbf{n}$ \footnote{
    This can be seen by using 
	$e^{A} B e^{-A}
	= \sum_{k=0} [A, B]_k /k!
	$, with $[A, B]_k$ the k-fold nested commutator, 
	and noting that all matrix elements $\mathbf{e}_i\cdot\hat{\mathsf{R}}\mathbf{e}_j$ commute.
    }.
	The imparted angular momentum thus  compensates the 
    orbital angular momentum of the emitted atom with respect to the nanoparticle center of mass.

	Finally, we note that Eq.~(\ref{eq:MeGenSgCh}) is closely related to the observable yielding the atom flux into the lab-frame direction $\mathbf{n}$.
    The latter is obtained from the correlated state of nanoparticle and atom by a partial trace over the atom Hilbert space,
	\begin{align}\label{eq:MassiveFlux}
		\hat{\Phi} (\mathbf{n})
		={}&
		\lim\limits_{r\to \infty} r^2 \mathbf{n} \cdot
        \tr_\mathrm{atom} [ \hat{D} \rho \otimes \proj{\psi_\downarrow} \hat{D}^\dagger \mathbf{j} (r \mathbf{n} )]
		\\={}&
		\lim\limits_{r\to \infty} r^2 \mathbf{n} \cdot \dfrac{1}{m}  \text{Re} \left[
		\braket{\psi_\downarrow | \hat{D}^\dagger \delta (r \mathbf{n} - \mathbf{x}) \mathbf{p} \hat{D} | \psi_{\downarrow}}
		\right]
        \nonumber
		,
	\end{align}
    where the first line involves the the atomic probability current density operator
    $\mathbf{j} (\mathbf{r}) = \{\delta(\mathbf{x} - \mathbf{r}), \mathbf{p} \}/2m$.
    Using the general initial state (\ref{eq:psidown}), we insert a position basis and utilize that the retarded Green function in an unbounded domain fulfills the Sommerfeld radiation condition $\lim_{r \to \infty} r (\partial_r - i p_0 /\hbar) G^+ (\mathbf{r}, \mathbf{s}, E_0) = 0$  \cite{Schot1992}
	to arrive at

	\begin{align}
		\hat{\Phi} (\mathbf{n}) ={}&
		\lim\limits_{r\to \infty} r^2
		\dfrac{\hbar^2 g^2 p_0}{m}
		\bra{\varphi_0} G^\dagger (E_0 + i \hbar \Gamma_0 /2) \hat{D}^\dagger
		\ket{r \mathbf{n}} \nonumber\\&\times \bra{r \mathbf{n}}
		\hat{D}
		G (E_0 + i \hbar \Gamma_0/2) \ket{\varphi_0}
        .
        \end{align}
        This limit can now be related to the momentum-position matrix elements of the M\o{}ller operators,
        as follows by identifying the amplitudes of emission (\ref{eq:EmAmDef}) and invoking
        (\ref{eq:MoMatEl}). Taking note of the definition (\ref{eq:DefGam0}) of $\Gamma_0$, the flux operator finally takes the form
        \begin{align}
        \hat{\Phi} (\mathbf{n})
		={}&
		\Gamma_0
		\dfrac{
			\left|
			\braket{p_0  \mathbf{n} | \hat{D} \Omega_-^\dagger | \varphi_0 }
			\right|^2}
		{\int_{S^2}\Diff2n' \,
			| \braket{p_0 \mathbf{n}' | \Omega_-^\dagger | \varphi_0 }|^2}
		.
	\end{align}
	Comparison with Eq.~(\ref{eq:defL}) shows that this expression is the modulus squared of the jump operators, as stated in Eq.~(\ref{eq:GenFlux}). It demonstrates that the flux operator is related to the flux introduced in Sect.~\ref{chap:Deso} by an operator-valued rotation,
    $\hat{\Phi} (\mathbf{n})=\Phi (\hat{\mathsf{R}}^T \mathbf{n})$.

    It is now straightforward to generalize the above treatment to 
 	the case of many independent emission sites. This amounts to taking a  sum of dissipators of the form (\ref{eq:MeGenSgCh}) for many atoms initially bound at positions $\mathbf{s}$ and energies $E$.  In the continuum limit, the jump operators (\ref{eq:defL}) then take the form (\ref{eq:LGen}), and the general master equation reads
\begin{align}
        \label{eq:MEGen}
		\mathcal{D} \rho ={}&
		\int\limits_0^\infty \diff E
		\int\limits_{ V} \Diff3s
		\int\limits_{S^2} \Diff 2 n\,
        \big[\hat{L}(\mathbf{n}, \mathbf{s}, E) \rho \hat{L}^\dagger(\mathbf{n}, \mathbf{s}, E)
		\nonumber\\&-  |\hat{L}(\mathbf{n}, \mathbf{s}, E)|^2 \rho \big].
	\end{align}
    In practice, the microscopic details of the emission process are typically unknown and only the atom flux density $\Phi (\mathbf{n}, \mathbf{s}, E)$ is available. 
    Even in this case, the essential form of the jump operators can  be specified by means of their polar decomposition.  Eq.~(\ref{eq:GenFlux}) fixes their positive part, while  it is clear 
    from the above  construction that the unitary must be of the form $e^{- i p \mathbf{n} \cdot \hat{\mathbf{X}}/ \hbar } \hat{U}_\mathrm{rot}$.
    In the absence of any further information, setting $\hat{U}_\mathrm{rot}=e^{- i p(E) \mathbf{n} \cdot \hat{\mathsf{R}} \mathbf{s}/\hbar }$ as in (\ref{eq:LDes}) may be considered the natural choice in that it describes precisely the recoil to the emitted atoms orbital angular momentum, as seen above in Eq.~(\ref{eq:A0def}).

\subsection{The localization rate for desorption}

    The jump operators (\ref{eq:LDes})
    and (\ref{eq:LGen})
    are diagonal in position-orientation basis, and the linear motion is only subject to momentum kicks.
    This means that the spatio-orientational representation of the master equation takes the simple form  $\braket{\mathbf{X}, \mathsf{R}| \mathcal{D} \rho | \mathbf{X}', \mathsf{R}'} = - F_{\mathsf{R}, \mathsf{R}'}(\mathbf{X}- \mathbf{X}') \braket{\mathbf{X}, \mathsf{R}| \rho | \mathbf{X}', \mathsf{R}'} $ specified by the complex rate $F_{\mathsf{R}, \mathsf{R}'}( \Delta \mathbf{X} )$.
    Its real part is non-negative, describing an exponential decay of spatio-orientational coherences. It is referred to as the localization rate, since the loss of coherence implies that a delocalized state turns into a mixture of localized ones over time.
     
    The localization rate can be simplified in the case of the desorption of atoms from the nanoparticle surface.  By using
    $2 a b^* - |a|^2 - |b|^2 = -|a-b|^2 + 2 i \, \text{Im}[ a b^*]$, it reads
    \begin{align}
        &\mathrm{Re} \, F_{\mathsf{R}, \mathsf{R}'}( \Delta \mathbf{X} )
		={}
        \int\limits_0^\infty \diff E \int\limits_{\partial V} \Diff 2 s \int\limits_{S^2} \Diff 2 n \, \dfrac{1}{2}
        \Big[
        \Phi( \mathsf{R}^T \mathbf{n}, \mathbf{s}, E)
        \nonumber\\&+ \Phi( \mathsf{R}^{\prime T} \mathbf{n}, \mathbf{s}, E)
        - 2\sqrt{\Phi( \mathsf{R}^T \mathbf{n}, \mathbf{s}, E) \Phi( \mathsf{R}^{\prime T} \mathbf{n}, \mathbf{s}, E)}
        \nonumber\\&\times
        \cos\Big(\dfrac{p(E)}{\hbar} \mathbf{n} \cdot [\Delta \mathbf{X} + (\mathsf{R} - \mathsf{R}') \mathbf{s}]\Big)
        \Big].
    \end{align}
    In this expression one can again identify the two different contributions to the decoherence process discussed in Sect.~\ref{chap:DesoME}.
    In the limit of negligible recoil, $p \to 0$, the integrand becomes $(\sqrt{\Phi_{\mathsf{R}} } - \sqrt
    {\Phi_{\mathsf{R}'}})^2 /2$, a measure of how discernible the two orientations are from the emitted flux density.     On the other hand, even if the flux density were approximately independent of the emission direction, a non-zero recoil would lead to decoherence.
    
    Finally, we note that the limit of an arbitrarily large recoil turns the decoherence rate into the total rate of atom emission because the cosine term becomes highly oscillatory. Moreover, the decoherence rate is bounded by twice the emission rate.
    Both of these statements also hold for the general outgassing master equation.
    To get a rough estimate for the strength of the effect, consider the specific outgassing rate $ 6.6\times 10^{-9} \, \mathrm{Pa} \, \mathrm{m}^3 / \mathrm{s} \, \mathrm{m}^2$ \cite{Battes2021}
    of bulk silica at room temperature $30 \, \mathrm{h}$ after baking.
    For a spherical nanoparticle with a diameter of of $150 \, \mathrm{nm}$ this yields a total emission rate of $0.33 \, \mathrm{Hz}$.
    The rate for untreated bulk gold at room temperature is of the order of $8.5\times 10^{-8} \, \mathrm{Torr} \, \mathrm{l} / \mathrm{cm}^2 \, \mathrm{s}$ \cite{Grinham2017}. Taking this at face value yields a total emission rate of about $2 \, \mathrm{kHz}$ for a particle of the same size.

    \section{Discussion and Conclusion}
    In this article, we investigated how the stochastic emission of small constituents affects the motional quantum state of a nanoparticle.
    The emission characteristics were obtained by considering the escape of a scalar particle metastably bound to the emitting body. This allowed us to define amplitudes of emission and relate them to the M\o{}ller operators from  scattering theory.
    To arrive at a general Markovian master equation (\ref{eq:MEGen}), we assumed the nanoparticle motion
    to be slow on the emission timescale, and the emission characteristics to be stationary on the scale of the decoherence dynamics.

    As an important property of the master equation, we found that the squared modulus of the jump operators can be identified with the emission flux density operator, see Eq.~(\ref{eq:GenFlux}). The latter can be readily obtained by promoting the emission rates from empirical data or phenomenological models to an operator.
    In principle, the jump operators (\ref{eq:LGen}) must be calculated from the  Green function associated with the potential imposed by the nanoparticle on the escaping particle, see Eq.~(\ref{eq:Adef}).
    Yet, their positive parts are fully determined by the emission flux density operator, see Eq.~(\ref{eq:GenFlux}), and we have argued that  their unitary parts can be approximated by the mechanically expected linear and angular momentum recoil associated with an emission event.

    For the case of the desorption of adsorbates from the particle surface, we presented a simplified master equation~(\ref{eq:MEDesoDiff}). It is valid in the case of small spatio-orientational delocalizations, as expected for large particles, describing diffusion and decoherence on equal footing. It is characterized by a matrix of momentum diffusion tensors and  a thermophoresis-like force and torque, which agree with classical results. 
	
	The presented quantum master equations facilitate accurate estimation of a decoherence processes which may become relevant in future quantum experiments in the field of levitated optomechanics. 
    A notable feature of their derivation is the use of M\o{}ller operators to non-perturbatively describe the interaction of an emitted particle with the emitting body during the former's escape. This approach may find use beyond the problem discussed here, e.g.\ if the emitted particles exhibit internal structure or intrinsic angular momentum. Moreover, we expect the  amplitudes of absorption defined in (\ref{eq:EmAmDef}), see also (\ref{eq:MoMatEl}), to permit describing decoherence due to absorptive processes,  such as resublimation or absorption of heat radiation.

    The presented treatment ceases to be valid for small emission energies, if the dwell time of the emitted particle becomes comparable to the motional time scale of the emitting body. In this case, one can no longer describe the particle degrees of freedom as entering parametrically through $\hat{D}$, but needs to calculate the M\o{}ller operators on the total nanoparticle-escaping particle Hilbert space. Moroever,  we assumed the emission to take place into vacuum, an approximation bound to fail if the nanoparticle is very close to a surface,  as might be the case in on-chip setups for levitation. How to include such structured environments into the presented approach is still an open problem.

	\section*{Acknowledgments}
	
	JS and KH acknowledge funding by the DFG --
	515993674.
	BAS is supported by the DFG -- 510794108 and by the Carl Zeiss foundation through the project QPhoton.

\end{document}